\input{aipcheck}

\documentclass[
    ,final            
  ]
  {aipproc}

\layoutstyle{8x11single}

\begin{document}

\title{Spectral Lags of Gamma-Ray Bursts from Primordial Black Hole (PBH) Evaporations}



\author{T. N. Ukwatta}{
  address={The George Washington University, Washington, D.C. 20052, USA}
  ,altaddress={NASA Goddard Space Flight Center, Greenbelt, MD 20771, USA}
}

\author{J. H. MacGibbon}{
  address={Department of Physics and Chemistry, University of North Florida, Jacksonville, FL 32224, USA}
}

\author{W. C. Parke}{
  address={The George Washington University, Washington, D.C. 20052, USA}
}

\author{K. S. Dhuga}{
  address={The George Washington University, Washington, D.C. 20052, USA}
}


\author{A. Eskandarian}{
  address={The George Washington University, Washington, D.C. 20052, USA}
}

\author{N. Gehrels}{
  address={NASA Goddard Space Flight Center, Greenbelt, MD 20771, USA}
}

\author{L. Maximon}{
  address={The George Washington University, Washington, D.C. 20052, USA}
}

\author{D. C. Morris}{
  address={The George Washington University, Washington, D.C. 20052, USA}
  ,altaddress={NASA Goddard Space Flight Center, Greenbelt, MD 20771, USA}
}


\begin{abstract}
Primordial Black Holes (PBHs), which may have been created in the
early Universe, are predicted to be detectable by their Hawking
radiation. PBHs with an initial mass of $\sim 5 \times 10^{14}
\rm\, g$ should be expiring today with a burst of high energy
particles. Evaporating PBHs in the solar neighborhood are
candidate Gamma-Ray Bursts (GRBs) progenitors. We propose spectral
lag, which is the temporal delay between the high energy photon
pulse and the low energy photon pulse, as a possible method to
detect PBH evaporation events with the Fermi Gamma-ray Space
Telescope Observatory.
\end{abstract}

\maketitle


\section{Introduction}
In the present era black holes with mass less than a few solar
masses are not expected to be created by any known process. Early
in the Universe, however, primordial black holes (PBHs) could have
been created with masses ranging from the Planck mass
($10^{-5}\,\rm g$) to as large as $10^5\,M_\odot$, or larger. PBH
formation scenarios include the collapse of overdense regions due
to primordial inhomogeneities, especially those generated by
inflation, a softening of the equation of state, or bubble
collisions at cosmological phase transitions, and the collapse of
oscillating cosmic string loops or domain walls. (For a recent
review see \cite{C05}).
\\ \\
Hawking discovered that, due to thermodynamical requirements and
quantum-gravitational effects, all black holes continually radiate
particles thermally \cite{H1,H2}. The Hawking temperature $T$ is
inversely proportional to the black hole mass $M$. Thus, as the
black hole radiates its temperature increases. It can be shown
that PBHs with an initial mass of $\sim 5.0 \times 10^{14} \rm{\,
g}$ should be expiring today in a burst of high energy particles
\cite{MCP}. In this context evaporating PBHs in the solar
neighborhood are candidate Gamma-Ray Bursts (GRBs) progenitors.

\section{Hawking Radiation}
According to quantum theory, virtual particles are continuously
created and destroyed in the vacuum. Heuristically, one way to
interpret the Hawking radiation process invokes the strong
gravitational field gradient near the event horizon of the black
hole. This field gradient can separate particle-antiparticle
pairs. In some cases, one particle falls with apparent negative
energy into the black hole, while the remaining one has sufficient
positive energy to escape to infinity. As a result, some particles
can come out of the vacuum as real particles by obtaining energy
from the black hole.
\\ \\
Black holes will predominantly radiate particles whose de Broglie
wavelength ($\lambda$) is roughly of the order of the
Schwarzschild radius of the black hole ($R_s \equiv 2GM/c^2$) and
so the energy of radiated particles can be estimated as follows
\cite{P1}
\begin{equation}
E =\frac{h c}{\lambda} \sim kT = \frac{\hbar c^3}{8\pi GM}.
\end{equation}
From the thermodynamic Stefan-Boltzmann relation, the Hawking
luminosity can be estimated by the area of the black hole horizon
times the radiation intensity (with $a$ proportional to the number
of degrees of freedom of the radiated particles) as follows:
\begin{equation}
L \equiv - \frac{dM}{dt} \, c^2 \approx (4 \pi R_s^2)(aT^4)
\propto \frac{1}{M^2}.
\end{equation}
Thus, the time remaining until the black hole expires by radiating
away its mass is roughly given by
\begin{equation}
t \propto M^{\,3}.
\end{equation}
The estimated black hole temperature, luminosity and time to
expire are summarized below
\begin{equation}
T \approx 10 \bigg[\frac{10^{15} \rm{g}}{M}\bigg]
\rm{MeV};\,\,\,\,\emph{L} \approx 10^{20} \bigg[\frac{10^{15}
\rm{g}}{\emph{M}}\bigg]^{2} \rm{erg \, s^{-1}};\,\,\,\, \emph{t}
\sim 10^{10} \bigg[\frac{\emph{M}}{10^{15} \rm{g}}\bigg]^3
\rm{yr}.
\end{equation}
The latter two relations are modified by the contributions of
various particle species producing Hawking radiation and
subsequent decay processes \cite{M2}.

\section{Photon Spectra from PBHs}
\begin{figure}[htp]
  \includegraphics[height=.28\textheight]{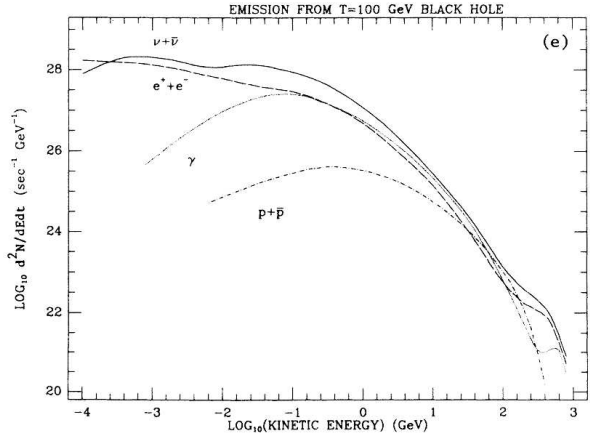}
  \caption{Instantaneous emission from a 100 GeV black hole~\cite{MW}.\label{Fig01}}
\end{figure}

A black hole should directly emit those particles which appear
non-composite compared to the wavelength of the radiated energy
(or equivalently the black hole size) at a given temperature. When
the temperature of the PBH exceeds the Quantum Chromodynamics
(QCD) confinement scale (250-300 MeV), quarks and gluons will be
emitted \cite{MW}. These particles should fragment and hadronize,
analogous to the jets seen in accelerators, as they stream away
from the black hole \cite{MCP}. The jets will decay on
astrophysical timescales into photons, neutrinos, electrons,
positrons, protons and anti-protons. The particle spectra of
direct as well as decay products are shown in Fig~\ref{Fig01} for
a $T=100$ GeV black hole. Most of the photon flux seen at
astrophysical distances are jet decay products.
\\ \\
The average energy of the photons ($\overline{E}_{\gamma}$)
emitted by the PBH scales, for black hole temperatures $T \sim 0.3
- 100$ GeV, as \cite{MW}
\begin{equation}
\overline{E}_{\gamma}\approx 3 \times 10^{-1}
\left(\frac{T}{\rm{GeV}}\right)^{0.5} \rm{GeV}
\end{equation}
Hence, most of the photons emitted by $T < 1$ TeV PBHs are in the
Fermi Large Area Telescope (LAT) energy range ($20 \rm{\,\, MeV} <
E < 300 \rm{\,\, GeV}$). The emitted flux of photons
($\dot{N}_{\gamma}$) scales for $T \sim 0.3 - 100$ GeV as
\cite{MW}
\begin{equation}
\dot{N}_{\gamma}\approx 2 \times 10^{24}
\left(\frac{T}{\rm{GeV}}\right)^{1.6} \rm{s^{-1}}.
\end{equation}
The time ($t$) left until the PBH completes its evaporation is
calculated in Table 1 for a range of black holes temperatures
using the method of reference \cite{M2}.
\\ \\
A PBH burst should produce $n$ photons in a detector of effective
area $A_{\rm eff}$ and angular resolution $\Omega$ if it is closer
than
\begin{equation}
d\simeq 0.03 n^{-0.5} \left(\frac{A_{\rm
eff}}{\rm{m^2}}\right)^{0.5}\left(\frac{T}{\rm{TeV}}\right)^{-0.7}
\rm pc
\end{equation}
and be detectable above the extragalactic gamma ray background at
energy $E$ if it is closer than
\begin{equation}
d\simeq 0.04
\left(\frac{\Omega}{\rm{sr}}\right)^{-0.5}\left(\frac{E}{\rm{GeV}}\right)^{0.7}\left(\frac{T}{\rm{TeV}}\right)^{0.8}
\rm pc.
\end{equation}
If PBHs are clustered in our galaxy with local density enhancement
factor $f_{\rm local}$ then the number presently expiring (i.e.
PBH evaporation rate $R$) is given by \cite{MC}
\begin{equation}
R\leq 10^{-7}\,f_{\rm local}\, \rm{pc^{-3} \, yr^{-1}}.
\end{equation}
A typical estimate for Galactic (halo) $f_{\rm local}$ is $\sim
10^6$ or larger \cite{MC}. Thus, such bursts may be observable
with the LAT. Conversely, non-detection of PBHs by the LAT may
lead to tighter bounds on the PBH distribution.
\begin{table}
\begin{tabular}{llllllll}
\hline
Model  & 1 Hour & 1 min & 30 sec & 10 sec & 1 sec & 30 ms & 10 ms  \\
\hline
Standard Model & 294 GeV & 1.15 TeV & 1.45 TeV & 2.09 TeV & 4.51 TeV & 14.5 TeV & 20.9 TeV \\
Higgs & 250 GeV & 0.98 TeV & 1.24 TeV & 1.8 TeV & 3.8 TeV & 12.0 TeV & 18.0 TeV \\
SUSY  & 206 GeV & 0.81 TeV & 1.0 TeV & 1.5 TeV & 3.2 TeV & 10.0 TeV & 15.0 TeV \\
\hline
\end{tabular}
\caption{Temperatures of the PBH according to various high energy
models when its remaining lifetime until total evaporation is 1
hour, 1 min, 30 sec, 10 sec, 1 sec, 30 ms and 10 ms, using the
method of \cite{M2}.} \label{tab:a}
\end{table}

\section{Discussion}
With the launch of the Fermi Gamma-ray Space Telescope observatory
on June 11, 2008, we have a new opportunity to examine the high
energy band pass which is suited to detect PBH evaporation events.
The discovery of a PBH would provide a unique probe of many areas
of physics including the early Universe, gravitational collapse
mechanisms, dark matter, and quantum gravity. Positive PBH burst
detection should also elucidate extensions of the Standard Model,
e.g. the existence of Higgs boson or Supersymmetry.
\\ \\
One approach to identify a PBH evaporation event directly is to
look at the spectrum spanning from MeV to GeV energy scales.
However, getting enough photons from these fast transient events
to make a spectrum can be difficult. In contrast, spectral lag
measurements merely require a light curve in two energy bands and
does not need many counts. Therefore we can measure the spectral
lag even for weak events that last for very short time scales.
Hence measuring spectral lag is a possible method to identify
PBHs. Qualitative analysis of the spectral lag of PBHs shows
positive to negative evolution with increasing energy. Work is in
progress to calculate quantitative values for the PBH spectral
lags, including the low energy inner bremsstrahlung contribution
\cite{PCM}.

\end{document}